\documentclass[aip,pop,amsmath,amssymb,superscriptaddress,twocolumn,nofootinbib,floatfix,showpacs]{revtex4}
\usepackage{bm,bbm,amsmath,amssymb,color,subeqnarray,graphicx,dcolumn}
\usepackage{graphicx,epsfig,latexsym,float,subfloat,subfigure,ulem,gensymb,varwidth,booktabs}
\usepackage{comment}
\usepackage{caption}
\usepackage{appendix}
\begin{document}
\title{ Compton Scattering in Plasma: Multiple Scattering Effects and Application to Laser-Plasma Acceleration}
\author{Ravindra Kumar}
\affiliation{Department of Physics, Indian Institute of Technology Kanpur, Kanpur-208016 (India)}
\author{V. Ravishankar}
\affiliation{Department of Physics, Indian Institute of Technology Kanpur, Kanpur-208016 (India)}
\affiliation{Department of Physics, Indian Institute of Technology Delhi, Hauz Khas, New Delhi-110016 (India)}

\begin{abstract}
\noindent We explore the physics of electron acceleration in a plasma medium in an effective field theory framework. 
Employing a multiple Compton scattering mechanism, it is found that the acceleration can be sustained in such a medium so as to attain the energies up to the order of $O(100~ \rm{MeV})$ within a centimeter. Also, the collimation and mono-energetic electron spectrum can be obtained by proper tuning of the plasma parameters with the photon frequency. The present work is potentially useful in understanding the physics of laser-plasma accelerators.
\end{abstract}
\pacs{ 41.75.Jv,52.38.Kd, 52.40.Mj,52.25.Mq} \maketitle
\section{ Introduction}\label{sec:intro}

\noindent The collective behavior of the plasma plays a vital role in laser plasma interactions. 
This has been a field of intense activity, spanning application to astrophysical phenomena\cite{chandrash}, material science\cite{faust1,faust2,soffel},  and Germanium telescope technology\cite{boggs,brain}. Of particular interest is the application to laser plasma accelerators (LPA), which holds the promise of producing quality beams of very high energy produced over very short distances \cite{dt}. When a laser of high power ($\sim  10^{18}~ \rm{W/cm^3}$) interacts with a high-density plasma (density $\sim 10^{18}~ \rm{cm^{-3}}$), various plasma waves are generated in the medium. The electric fields associated with such waves create an accelerating gradient of the order $O(1~\rm{GeV/cm}$) which is predominantly in a longitudinal direction. If some electrons either from the plasma background or externally injected are trapped in a proper phase with such fields, they may be accelerated to a relativistically high energy in a very short distance. Thus various particle accelerators has been proposed (see Refs. \cite{joshi1, esarey} for a full discussion). 
           
An effective field theoretic (EFT) description of Compton scattering has been recently developed in Refs. \cite{ravi1, ravi2} (hereafter referred to as I and II); Compton scattering of radiation with a plasma medium was studied where the collective behaviour of medium played a vital role.  Remarkably, it was found that the scattered electron from the plasma exhibits (i) an enhanced cross-section, (ii) a high degree of collimation and  (iii) a strong quasi-monochromatic behaviour, in some regions of the parameter space spanned by the plasma density and temperature.  Further, the strength of the equivalent accelerating field could be estimated to be $\sim 100~ \rm{MeV/cm}$. These features seem to bear a connection to the electron spectrum obtained in LPA, and it would be interesting to examine if the EFT description can be extended to understand the physics behind LPA.

The EFT analysis in I and II has several missing ingredients. (i) The stability of collimation and monochromaticity against multiple scatterings needs to be established. (ii) While the effective properties of the radiation were captured in the permittivity tensor, the modification of the electron properties, due to its interaction with the radiation, cannot be ignored. The modification can be incorporated by considering, not free electrons, but the solutions  that emerge from  Volkov equations \cite{volkov, berestetzkii}. Pardy has studied the solutions \cite{pardy4, pardy5} for some simple media. (iii) It  is  necessary to take into account the non-linearity in the dispersion relations for the photon. Finally, the effects of pulse shaping which is central to LPA is also not considered here. 

As the first step in this direction, we examine the stability of collimation and monochromaticity of the electron spectrum due to repeated scatterings. Moreover, we examine whether the electrons continue to gain energy with repeated scatterings, and whether the value can be expected to be anywhere near what is observed experimentally. That this task can be daunting can be appreciated by the fact that given the experimental parameters, the electron traverses a distance $d \approx 1~ \rm{nm}$ and gains, a few $ \rm{eV}$ in a single scattering, which is a minuscule fraction of the energy $\sim 100~ \rm{MeV-1GeV}$ attained over a distance of about $1~ \rm{cm}$. Furthermore, since we have completely ignored the all crucial pulse shape of the laser, it is not of much use to attempt anything more quantitative. If the approach fails even this test, any further attempt to model LPA in terms of an EFT would be futile. In this study, we do find that multiple scattering can impart the desired energy, indicating that the EFT description can be robust. It is worthwhile remembering at this stage that the earlier theoretical approaches have been computation intensive involving extensive numerical simulations based on the particle-in-cell approach \cite{lin, dawson, pukhov}, as may be seen, e.g, in Refs. \cite{malka, faure}.

            The paper has been organized as follows. In Sec. \ref{sec:sec_2}, we  review and summarize briefly the basic formalism that was developed in I and II. In Sec. \ref{sec:sec_3}, we  set up the basic algorithm for multiple-scattering computations and study the evolution of relevant observables  with the scatterings. Since it would require, for an electron, a large number of scatterings ($\sim N=10^8$) to gain the energy of order $O(100~\rm{MeV})$ and, it is a very difficult task to perform the computation for $N = O(10^8)$ scatterings; therefore, in order to get the electron spectrum after $N\sim 10^8$, we derive, in Sec. \ref{sec:sec_4}, an approximate formula for the electron energy and angle in the large $N$ limit. Then, the sensitivity of this (extrapolated) spectrum to the variation of the plasma parameters and the nature of the electron distribution is studied in Sec. \ref{sec:sec_5}. In Sec. \ref{sec:sec_6}, we shall point out the limitations of the present work and finally conclude the paper. 

\section{ Review of the Formalism}
\label{sec:sec_2}
\noindent  When a laser interacts with the plasma medium, various electromagnetic waves (also called plasma waves) are generated, which are basically of two kinds: transverse and longitudinal. Thus the associated Compton scattering involves both  transverse and longitudinal photons. These photons obey modified dispersion relations due to the permittivity of the medium. Thus the scattering process gets modified at all levels including the propagator and vertex and wave-function renormalization. We may, therefore, expect Compton scattering in the medium to differ significantly from free space scattering, both qualitatively and quantitatively.

\begin{figure}
  \begin{center}
    \includegraphics[clip,width= 3.5 in, height=8 cm]{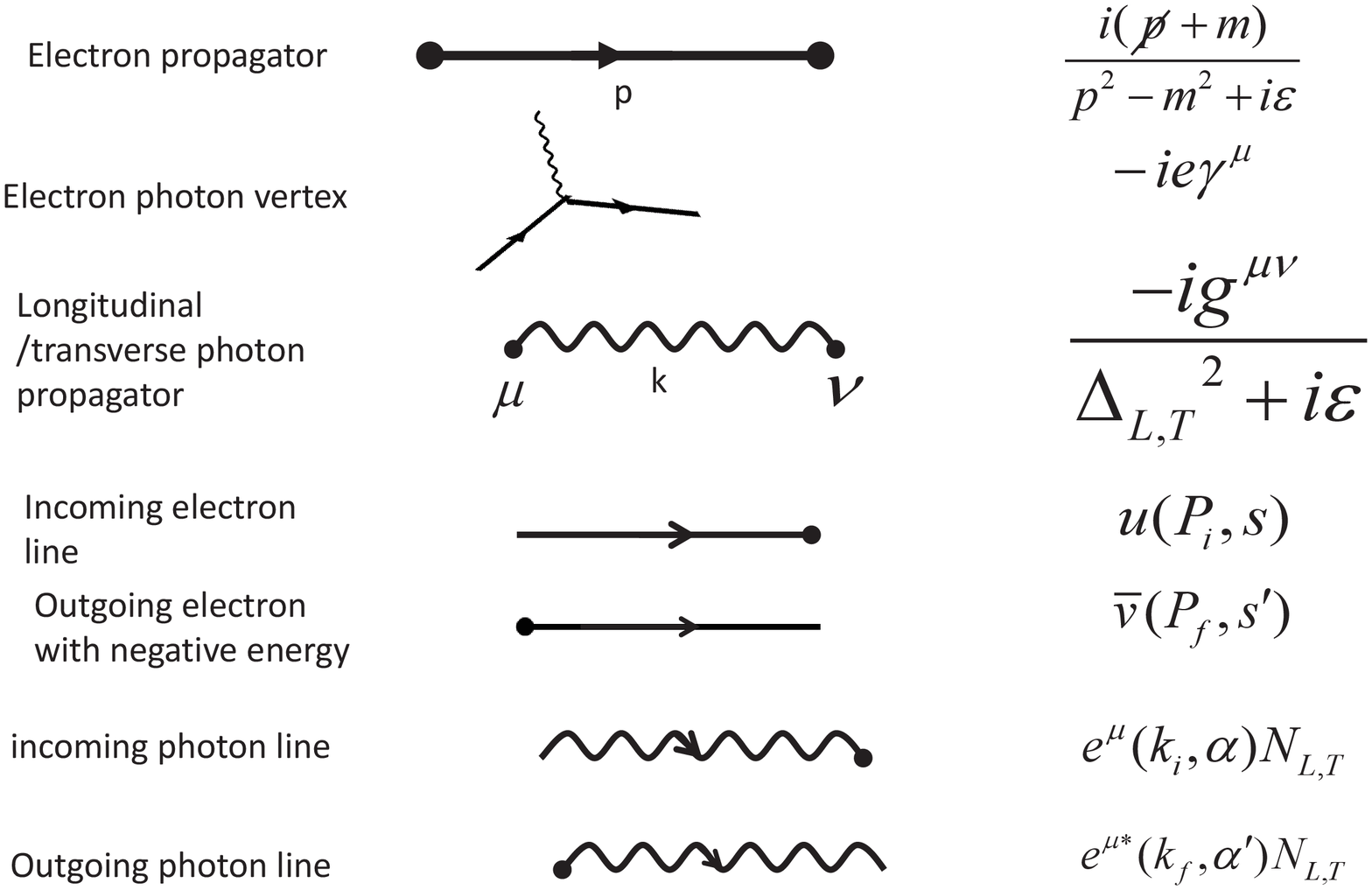}
    \caption{ The Feynman rules}
 \label{fey_rule3}
  \end{center}
\end{figure}

\begin{figure}
       \begin{center}
         \includegraphics[clip,width= 3.0 in, height=1.5 in]{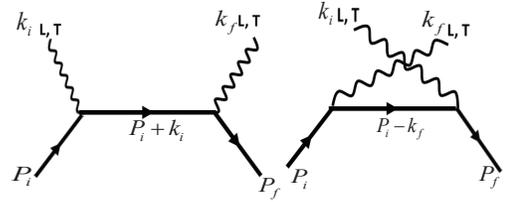}
         \caption{  Tree diagram
     for the Compton scattering. Here, $L$ and $T$ denote the longitudinal and transverse plasma-waves respectively.}
      \label{fey_daig2}
       \end{center}
\end{figure}

An EFT for this process has been proposed in I and II where all the above mentioned modifications were incorporated. For simplicity, only a linear dispersion relation for the virtual photons is considered. The corresponding Feynman rules and the Feynman diagram for the scattering are given, respectively, in Figs. \ref{fey_rule3} and \ref{fey_daig2}. We note that we employed the natural units $\hbar=c=k_B=1$ there and shall do the same throughout this paper.
 
The effective Lagrangian is given by 

  \begin{eqnarray}
   \mathcal{ L} =   \mathcal{L}_{Dirac}+\mathcal{L}_{Int}+\mathcal{L}_{Field},
  \end{eqnarray} 
   
 \noindent where $\mathcal{ L}_{Dirac} = \bar \psi (x)(i\gamma ^\mu  \partial _\mu   - m)\psi (x); ~
    \mathcal{ L}_{Int} = -e \bar \psi (x)\gamma ^\mu  \psi (x)A_\mu (x)$ 
    retain their usual forms while the field part gets modified to   
   \begin{equation}
   \mathcal{ L}_{Field} = \frac{1}{2}[{\cal P}^l_{ij}E_i {D}^l_j +
    {\cal P}^t_{ij} E_i {D}^t_{j}
   -{\vec B^2}], \label{efflag}
   \end{equation}
  where, $ {D}^{l,t}_i(\vec r,t)=\int{d^3{r^\prime} dt^\prime}{{\varepsilon}^{l,t}(t-t^\prime,\vec r-\vec r^\prime) E_i(t^\prime,r^\prime)}$, and $\varepsilon^{l,t}$ are the plasma (longitudinal and transverse) permittivities in the linear regime. ${\cal P}^{t,l}$ are respectively the projectors for the transverse and longitudinal modes. If we expand the photon wave functions in the standard plane wave basis, the transverse and the longitudinal modes suffer renormalizations because of the medium. The new renormalization factors  are given by \cite{harris} 

 \begin{eqnarray}
   {N_l}^{-1}=\sqrt{ \frac {k^2\partial(\omega {\varepsilon _l})}{\omega\partial\omega}}
   |_{\omega=\omega _{\vec k}},\,
    {N_t}^{-1}=\sqrt{\frac{\partial({\omega ^2}{\varepsilon _t})}{\partial\omega
   }}|_{\omega=\omega _{\vec k}}.
   \label{norm}
   \end{eqnarray}

We are, of course, considering the process 
\begin{equation}
\gamma(\omega_i,\vec{ k}_i, \alpha) + e^-(E_i,\vec{P}_i, s) \rightarrow \gamma(\omega_f, \vec{ k}_f, \alpha^{\prime}) + e^-(E_f, \vec{P}_f, s^{\prime}),
\label{process}
\end{equation}
where the energy, momentum and  spin variables are explicitly indicated for each particle.
For the dispersion relations between $\omega$ and $k$, which get determined by the permittivity of the medium, we assume linearity but go beyond the Gross-Bohm expression. Instead, we employ the Fried-Conte relations which are tabulated in \cite{conte}. This places a restriction on the allowed values of the photon energy when the mode is longitudinal: the ratio $g \equiv \omega_i/\omega_p \in [1,1.28]$, where $\omega_p$ is the  plasma frequency. There is no such restriction on the transverse mode. The kinematics is now fairly straightforward to work out, and it has been discussed in detail in Appendix \ref{app:appendix_b} of II.

We may express the cross-section for each choice of the photon polarization in the form

   \begin{eqnarray}
       d{\sigma_{\alpha\alpha^\prime}} = \frac{ N_\alpha^2N_{\alpha^\prime}^2d^3 P_f d^3 k_f
       }{16\pi^2E_i E_fv_{rel} }\delta^4(P_i  + k_i -
       P_f - k_f  )\left| {\bar M}_{\alpha\alpha^\prime} \right|^2,
       \label{crosssection}    
       \end{eqnarray}
where $\alpha,\alpha^{\prime} = L,~ T$ refer to photon polarizations in the initial and final states \cite{ravi2}.
The  form of  $\left| {\bar M}_{\alpha\alpha^\prime} \right|^2$ (note that we have summed over the electron spin) is 
considerably more involved than the Klein-Nishina formula. The complete expression may be found in the Appendix \ref{app:appendix_a} of this paper. Finally, we note that since we are dealing with a plasma at a temperature $T$, a further thermal average is required. All results of I and II which are  quoted in this section, were obtained after the thermal average. The cross-section is evaluated at the tree level (see the diagram in Fig. \ref{fey_daig2}).

\subsection{Summary of the previous results}  
\noindent We briefly summarize the results obtained in I and II for continuity and to lay ground for discussion here. The initial photon energy was pegged at 
$\omega_i = 0.11~ \rm{eV}$ , and the plasma density varied in the range $5.36 \le n_e \le 8.53 \times 10^{19}~ \rm{{cm}^{-3}}$. The plasma was taken to be at three different temperatures  $T= 30,~ 50~ \rm{and} ~ 70~  \rm{eV}$. The results obtained can be broadly summarized as follows:
\begin{enumerate}
\item  The cross-section is completely dominated by the longitudinal plasmon mode, in particular, by $\sigma _{LL}~ \rm{and}~ \sigma_{TL}$; $\sigma_{LL}  \sim O(10^{12})$ and  $\sigma_{TL}  \sim O(10^{10})$  are larger than the other two contributions which are not appreciably different from the Klein -Nishina magnitudes. This result is valid over a region in the density-temperature plane. 
\item In the same region in the parameter space, the scattered electron spectrum reveals a high degree of collimation, with the angular dispersion $\Delta \theta \approx 6 ~\rm{mrad}$, and is also quasi-monochromatic, with $\Delta E/ E \approx 0.01$. Thus the scattered electron energy distribution is highly non-Maxwellian.
\item The scattering time (or, equivalently, the distance over which the electron gets scattered) can be estimated simply by using energy-time uncertainty argument. The effective accelerating electric field is estimated to be of $ O(100~ \rm{MV/cm})$ which is in accordance with the results obtained by simulations.
\item These remarkable features are, of course, not valid everywhere in the parameter space. Just as experimentalists observe, we find that when we move away from that region, collimation and momochromaticity both suffer deterioration in quality rapidly (e.g., see Ref. \cite{faure}). This can be quantified by a convenient parameter, the beam quality index defined by
\begin{equation}
Q = \Big( \Delta \theta \times \frac{\Delta E}{E} \times \frac{1}{f} \Big)^{-1},
\label{BQI}
\end{equation}
where $f$ is the fractional cross-section in the collimated region: $f= \Delta\sigma_{coll}/ \sigma_T$.
This parameter does a good job of capturing the essential qualities of the scattered electrons (see Fig. 6 of II, where the beam quality profile for varying densities and temperatures is shown). 

\end{enumerate}

The results summarized above naturally lead to the question as to their veracity when other contributions to laser plasma scattering which we listed in Sec. \ref{sec:intro} are incorporated. As mentioned, we look at one important component, namely, the multiple-scattering effects in the next section.

\section{Multiple-scattering Contributions} 
\label{sec:sec_3}

\subsection{The Basic Algorithm}
\noindent We first outline the algorithm which we employ to compute cross-sections after repeated scatterings.
(i) To avoid unnecessary tediousness, we drop the contributions from $TT$ and $LT$ scatterings both of which are minuscule. (ii) Since our primary interest is in the possible energy gain by the electron, we look at only those angles where the scattered electron gains the energy after each collision. 

\subsubsection{Kinematics of Scattering}

\noindent It has been shown in II that  the expressions for the scattering angles $\theta_f$ (for the photon) and $\theta_e$ (for the electron), both with respect to the direction of initial photon momentum $\vec k_i$, are given by  
   \begin{eqnarray}
         \theta_f & =& \cos^{-1}\big (\frac{C}{R}\big )+ \phi \label{theta_f}; \nonumber \\
         \theta_e & = & \cos^{-1}\big\{\frac{P_i\cos\theta_i+k_i-k_f\cos\theta_f}{P_f}\big\}
         \label{theta_e}
         \end{eqnarray}\\
          where $\theta_i$ is angle between the electron initial momentum $\vec P_i$ and the photon initial momentum  $\vec k_i$. The quantities that appear in Eq. (\ref{theta_e}) are given by 
             \begin{eqnarray}
         C & =&   \frac{A^2-E_f^2}{2k_f}; ~ R=\sqrt{a^2+b^2}; \nonumber \\
        \phi & =\  &\tan^{-1}(\frac{b}{a}); ~  a=k_i+P_i \cos\theta_i \nonumber \\
         A^2 &  = & E_i^2+k_i^2+k_f^2+2P_ik_i \cos \theta_i; 
         ~ b=P_i \sin \theta_i.
         \end{eqnarray} 

Thus the  algorithm can be easily expressed as a recursive equation
\begin{equation}
P_i^{(N+1)}\cos \theta_i^{(N+1)} =P_i^{(N)}\cos\theta_i^{(N)}+ k_i-k_f\cos\theta_f,
\label{recursive}
\end{equation}
where $\theta_i^{(N)}$  and  $\theta_i^{(N+1)}$ are, respectively, the electron scattering  angles before and after the $N^{th}$ scattering, and $P_i^{(N)}$ and $P_i^{(N+1)}$ are the magnitudes of respective momenta.  
  
It was seen in II that the differential cross-sections and the energy spectrum of the electrons were similar in the $LL$ and $TL$ cases, with the former being larger. However, a more careful look at the kinematics reveals that the energy gain, if there is to be any, will be dominated by  $\sigma_{TL}$. The reasons are two fold: (i) The restriction that $\omega/\omega_p \in [1,1.28]$ comes from the condition $\varepsilon_l(\omega, \vec{k})=0$ on the longitudinal photons.  Simple energetics implies that energy transfer in $LL$ scattering cannot exceed $0.28 \omega_p$.  However, even at $\omega_i = 20 \omega_p$ which we consider here, although $LL$ scattering is forbidden, $TL$ scattering is not only allowed, but can transfer a maximum energy $\Delta E \approx 18.7 \omega_p$ which is more than twenty five times of what is allowed in the $LL$ case. (ii) Interestingly, even in the allowed range, kinematics places further restrictions on energy transfer. We discuss this in Appendix \ref{app:appendix_b} of this paper.
Finally, since it serves no purpose to present the full electron spectrum after each scattering, we merely depict the evolution along the path of maximum probability, i.e., the angle and energy at which the cross-section peaks. In doing so, not much information is lost because of a high degree of collimation. In fact, since $\omega_f/\omega_p \in[1,1.28]$, the angular divergence is less than a milliradian.

\subsection{An Illustrative Example}
\noindent We choose the electron initial energy  $E_i=2m$, and the initial photon with frequency as $\omega_i=20\omega_p$, with its direction along the $Z$-axis. We choose the plasma parameters $n_e=6.02\times 10^{18}~ \rm{cm^{-3}}$ and $T=50~\rm{eV}$. These laser and plasma parameters values are very close to the experimental values (e.g., those in Ref. \cite{faure}). Further, for purposes of illustration, we choose the initial electron direction $\theta_i=\pi/2$. The general conclusions which we draw do not, of course, rely on this special choice.
 
\subsubsection{ Evolution of  Electron Energy and Momentum}

\begin{figure}
  \includegraphics[clip,width=7.0 cm,height=6 cm]{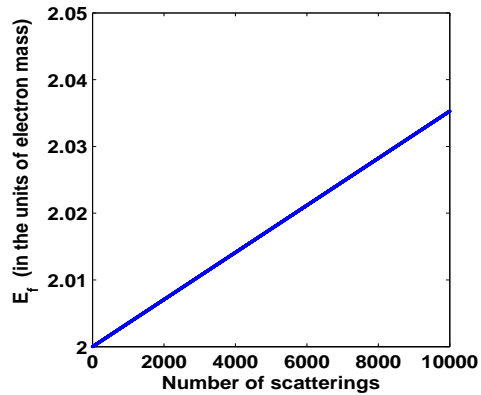}
   \caption{The electron energy with successive scatterings.}
  \label{Ef_multi}
  \end{figure}

\begin{figure}
 \includegraphics[clip,width=7.0 cm,height=6 cm]{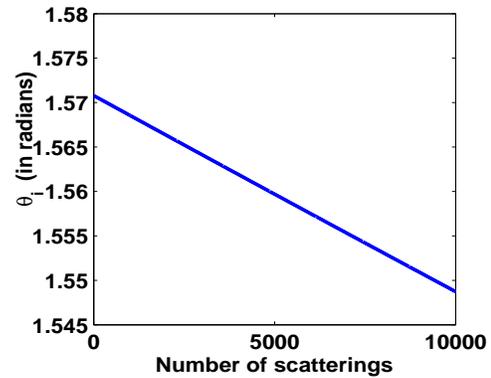}
  \caption{The electron scattering angle $\theta_e$ with successive scatterings.}
 \label{scat_angle_multi}
 \end{figure}

\noindent  Figs. \ref{Ef_multi} and  \ref{scat_angle_multi} show, respectively, the energy and the direction of the electron with the successive scatterings upto  $N=10^4$ scatterings. We see that the electron energy increases with subsequent scatterings, while its direction undergoes a slight tilt towards the initial photon direction.
 
\begin{figure}
 \includegraphics[clip,width=7.0 cm,height=6 cm]{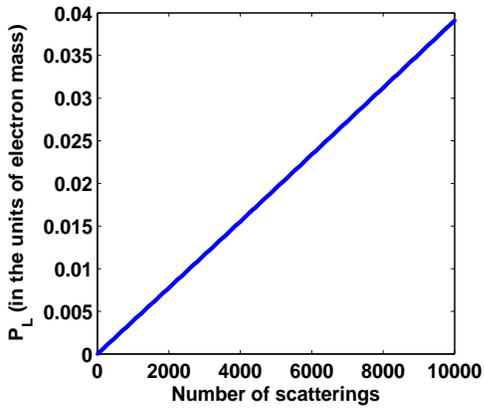}
  \caption{ Evolution of longitudinal momentum with successive scatterings. }
 \label{pl_multi}
 \end{figure}

\begin{figure}
 \includegraphics[clip,width=7.0 cm,height=6 cm]{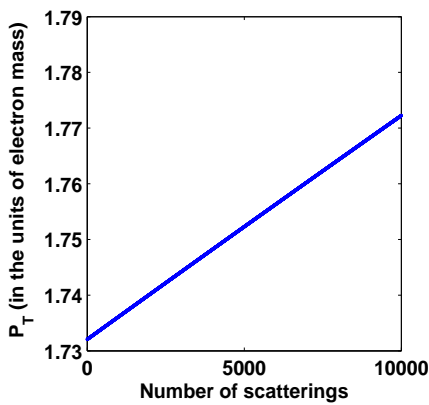}
  \caption{ Evolution of transverse momentum with successive scatterings.}
 \label{pt_multi}
 \end{figure}

  The evolution of longitudinal and transverse components of the electron momentum  are shown, respectively, in Figs. \ref{pl_multi} and  \ref{pt_multi}. The figures show almost a linear increase in both the components with  successive scatterings. 
 
\subsubsection {The Interaction Length}
\begin{figure}
\includegraphics[clip,width=8.0 cm,height=7 cm]{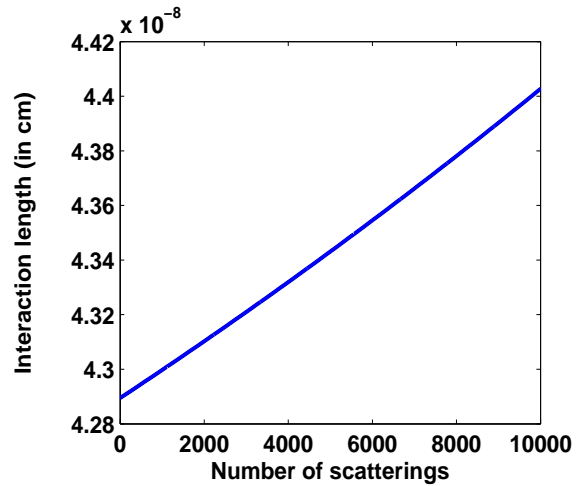}
 \caption{ The variation of the Compton interaction lengths with successive scatterings.}
\label{int_length}
\end{figure}

 \noindent The interaction time, or, equivalently, the interaction length, which is the distance travelled by the electron during the interaction along the path of maximum probability, is estimated using the uncertainty relation, as in I and II. This is depicted in Fig. \ref{int_length} which shows the interaction lengths involved during each scattering; they are all of $O(10^{-8}~ \rm{cm})$. Thus the total distance covered by the electron after $N=10^4$ scatterings is $\sim 1~ \rm{\mu m}$, with a corresponding energy gain of $\sim 0.02~\rm{MeV}$, as may be inferred from Fig. \ref{Ef_multi}. Since the increase in the energy is almost linear, and the order of the interaction length does not change much with subsequent scatterings, it is plausible that an energy of order $O(100~ \rm{MeV})$ can be gained by an electron within a traversal distance of a centimeter length. 

\subsubsection{Electronic Spectrum}

\begin{figure}
  \includegraphics[clip,width=9.0 cm,height=15 cm]{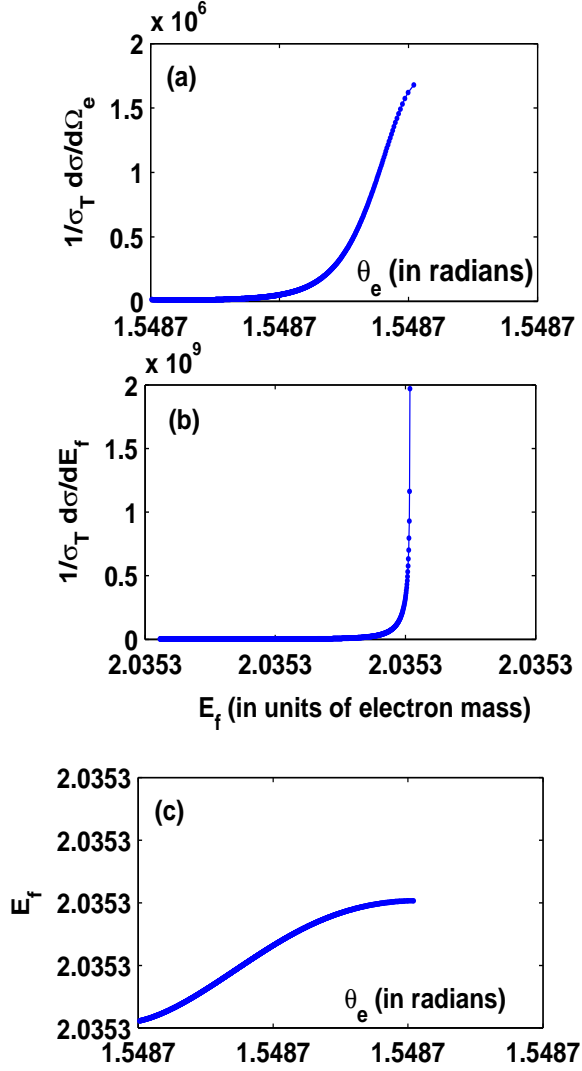}
   \caption{ Electron spectrum after $10^4$ scatterings}
  \label{prob_ef_theta}
  \end{figure}

\noindent We now turn our attention to the electron spectrum, i.e., $ \frac{1}{\sigma_T}\frac{d\sigma}{d\Omega_e}$ and  $\frac{1}{\sigma_T}\frac{d\sigma}{dE_f}$,  after $N=10^4$ scatterings which are shown, respectively, in Figs. \ref{prob_ef_theta}(a)  and  \ref{prob_ef_theta}(b). Fig. \ref{prob_ef_theta}(c) shows the energies of the scattered electron at various scattering angles. 
We can see that the whole spectrum is confined within one tenth of a milli-radian, with the energy spread, $\Delta E/E < 10^{-3} \%$. It is found that the energy and angular spread do not vary much with the number of scatterings.
 
Finally, we notice that both energy and momenta show a linear increase with scattering. It is necessary to dispel the possibility that these results are an artefact of the choice of initial electron angle, if further progress is to be made. We show the increase in electron energy with multiple scatterings for various initial angles, in Fig. \ref{ef_zeta_angle}(a). It is clear that the conclusions drawn above is not specific to the particular choice.

\section{Scattering with a Distribution of Electrons }
\label{sec:sec_4}
 \begin{figure}
     \includegraphics[clip,width=9.0 cm,height=11 cm]{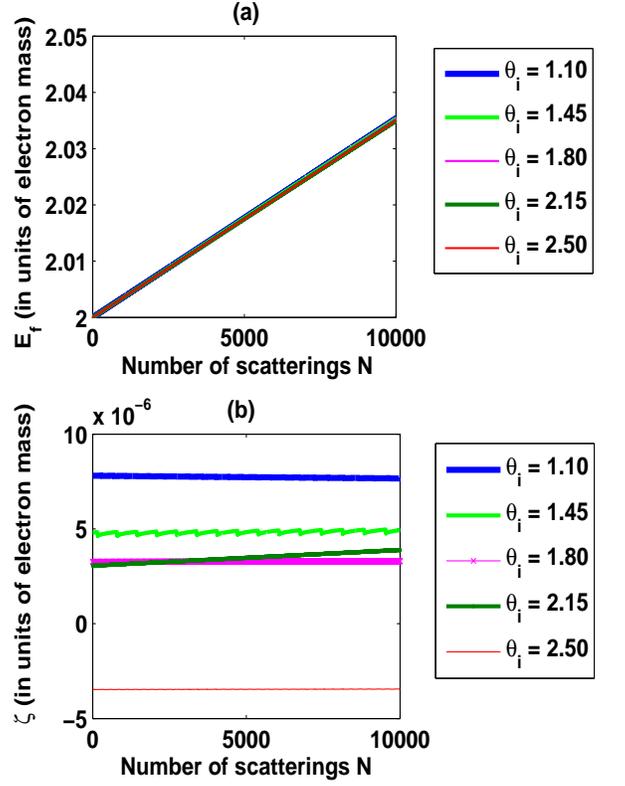}
      \caption{(a) The electron energy  and (b) the value of $\zeta$, defined in Eq. (\ref{zeta}), vs. number of scatterings N, for various initial angles. }
     \label{ef_zeta_angle}
     \end{figure}  
\noindent We have seen in the previous section that, for a specific choice of the initial electron direction, the angular and energy spread in the scattering almost remain unaltered with acceleration. Also, we found that the energy gain per scattering was independent of the initial electron angle, as depicted in Fig. \ref{ef_zeta_angle} (a). Using the independence, we develop here an approximation technique that allows us to estimate the final cross-section after a large number of scatterings.

At this stage, it is convenient to express energy in units of electron mass, in addition to setting $\hbar=c=k_B=1$. Let the energy transfer $E_f-E_i(= \omega_i-\omega_f \equiv \Delta \omega) =\Delta \omega_0$. Thus, after  $N$ scatterings, we get $ E_{N}=E_0 +N \Delta \omega_0$. Thus, making use of the  energy-momentum relation $E^2 = P^2 +1$, we obtain
    \begin{equation}
    P_N=\sqrt{P_0^2+2N\Delta\omega_0 E_0 + N^2{\Delta\omega_0}^2},
    \label{P_N}
    \end{equation}
which yields an approximate expression for the magnitude of the momentum.
   \begin{figure}
     \includegraphics[clip,width=8.0 cm,height=10 cm]{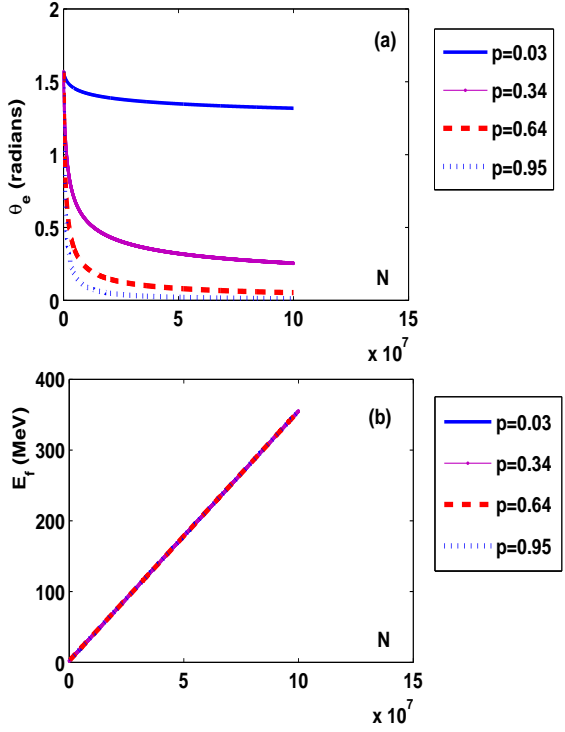}
     \caption{The extrapolated curves for (a) the electron angle $\theta_e$, and (b) the electron energy $E_f$, with respect to $N$, for different values of the exponent $p=\frac{\zeta_0}{2\Delta\omega_0}$ in Eq. (\ref {theta_N}).}
     \label{theta_e_Ef_vs_N}
     \end{figure}

     We now consider the scattering angle for the electron. First of all, the observation made in the illustration, that the scattering angle decreases with repeated scatterings, is borne out by Eq. (\ref{theta_e}), whenever there is a gain in $P_L$, i.e., $\Delta_l \equiv k_i -k_f \cos\theta_f > 0$.
   The momentum conservation equation, $\vec P_f =\vec P_i + \vec{k_i}-\vec{k_f} $, together with the fact that  $|{\Delta \vec K}|\equiv |{\vec k_i-\vec k_f}| \ll P_i$,  leads to the approximate expression
    \begin{equation}
   P_f \approx P_i +\Delta_l +k_i(\cos\theta_i-1) + O((\Delta K)^2).
    \label{pitopf_ratio}
    \end{equation}  
   Using the above approximation in Eq. (\ref{theta_e}), we get
    \begin{eqnarray}
   \cos \theta_e-\cos\theta_i&\approx &  (1-\cos\theta_i)\big( \frac{\zeta}{P_i} \big); \label{diff_theta_e_i} \nonumber  \\ \zeta & \equiv & (\Delta_l+k_i\cos\theta_i), 
   \label{zeta}
    \end{eqnarray}       
    which can now be written as the recursive relation
    \begin{eqnarray}
    \cos\theta_{N}-\cos\theta_{N-1}=(1-\cos\theta_{N-1})\frac{\zeta}{P_{N}}.
    \label{cos_recursive}
    \end{eqnarray}
   
    Remember that though $\zeta$ is a complicated function of ($\theta_i,~ P_i$), however, $\zeta \sim O(10^{-6}):~ \zeta  \ll P_i$ and therefore its dependence on the kinematical variables may be ignored as is also illustrated in Fig. \ref{ef_zeta_angle}(b). Therefore, considering a particular  value of $\zeta=\zeta_0$, corresponding to some initial values ($P_i=P_{i0}$, $\theta_i=\theta_{i0}$) and  $\Delta\omega=\Delta\omega_0$),  Eq. (\ref{cos_recursive}) attains the  following integral form:
    \begin{eqnarray}
    \int\limits_{\theta_0}^{\theta_N} \frac{d(\cos\theta)}{1-\cos\theta}=-\zeta_0\int\limits_{0}^{N}\frac{dx}{\sqrt{(E_0+x\Delta\omega_0)^2-1}},
    \label{integration}
    \end{eqnarray}
    where we have used momentum-energy relation apart from the linearity of electron energy with the number of scatterings. 

After performing the integration on both sides of Eq. (\ref{integration}), we obtain
    \begin{eqnarray}
    \theta_{N}=2\sin^{-1}\Big[\sin({\frac{\theta_0}{2}})\Big\{\frac{E_0+\sqrt{E_0^2+1}}{E_N+\sqrt{E_N^2+1}}\Big\}^p\Big],
    \label{theta_N}
    \end{eqnarray}
where the exponent $p \equiv\frac{\zeta_0}{2\Delta\omega_0}$. The decreasing behavior of the scattering angle, which was seen in Fig. \ref{scat_angle_multi}, can also be seen from Eq. (\ref{theta_N}). In other words, the electrons tend to align themselves towards the laser beam direction, with subsequent scatterings. 
 
 Figs. \ref{theta_e_Ef_vs_N} (a) and  \ref{theta_e_Ef_vs_N} (b) show, respectively, the extrapolated curves for the electron angle $\theta_e$ and the energy $E_f$ with $N$, for different the choices of the exponents $p$ which correspond to different initial conditions.
The curves for the electron angle, as in Fig. \ref{theta_e_Ef_vs_N} (a), are almost linear upto $N=10^4$, which is same as found by the actual computation (see Fig. \ref{scat_angle_multi}). The corresponding curves for the electron energy are shown in the Fig. \ref{theta_e_Ef_vs_N} (b), which are almost in complete overlap with the electron energy curve found by the actual computation and depicted in Fig. \ref{Ef_multi}.
 \begin{figure}
   \includegraphics[clip,width=8.0 cm,height=11 cm]{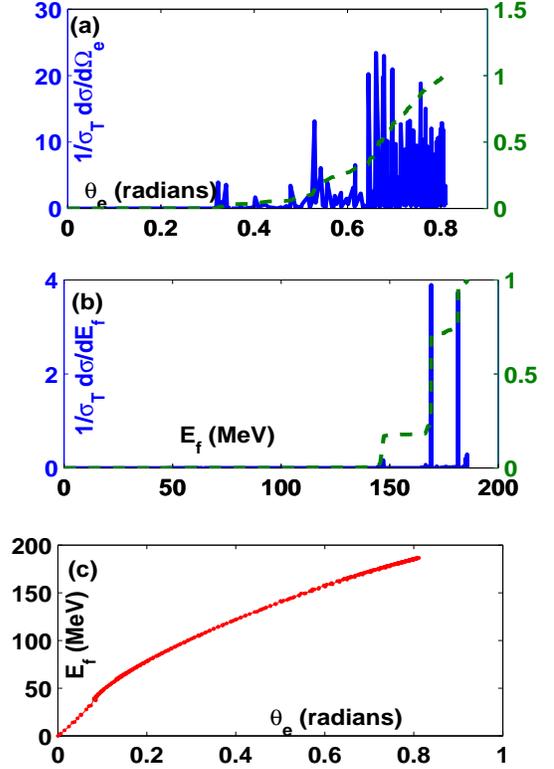}
   \caption{ The electron spectrum after $N=10^8$ scatterings with $\omega_i=20\omega_p$, $n_e =6.02\times 10^{19}~ \rm{cm^{-3}}$ and $T=50~\rm{eV}$, when with initial electron distribution confined in the angular region $\theta_i\in[2.1,2.5]$. The beam quality index $Q=3.58$.}
   \label{spectrum20}
   \end{figure} 
 \begin{figure}
   \includegraphics[clip,width=8.0 cm,height=11 cm]{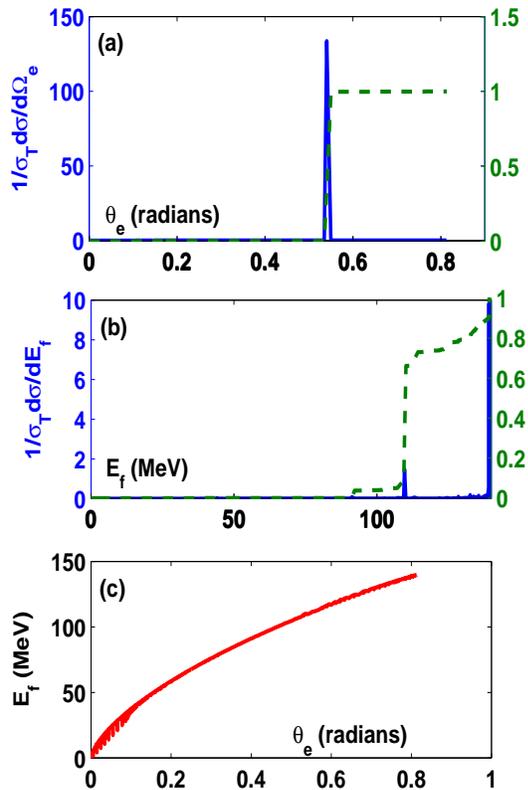}
   \caption{The electron spectrum after $N=10^8$ scatterings with $\omega_i=16\omega_p$, $n_e =5.53\times 10^{19}~ \rm{cm^{-3}}$ and $T=50~\rm{eV}$, when the initial electron distribution confined in the angular region $\theta_i\in[2.1,2.5]$. The beam quality index $Q=48.33$.}
   \label{spectrum}
   \end{figure} 

We now present results for the electron spectrum $\frac{1}{\sigma_T}\frac{d\sigma}{d\Omega_e}$, $\frac{1}{\sigma_T}\frac{d\sigma}{dE_f}$ after $N=10^8$ scatterings. Using the above extrapolated formula, in the following, we illustrate the acceleration of a bunch of electrons. Importantly, we assume that the whole distribution should be within a distance of a plasma wavelength ($\sim$ 10 microns). The final electron spectrum is found to be sensitive on initial frequency, nature of the electron distribution and plasma density.

 For the sake of illustration, let us  choose a  distribution of electrons which is a Gaussian in energy having a peak at the energy $1~\rm{MeV}$ and with $4\%$ of energy spread, however, they are uniformly distributed in the angular region within $\theta_i \in [2.1,2.5]$. We again choose the initial plasma frequency to be $\omega_i=20\omega_p$ with plasma density $6.02\times 10^{18}~\rm{cm^{-3}}$ and temperature $T=50~\rm{eV}$. The final spectrum is shown in  Fig. \ref{spectrum20}. The electron angular and energy spectra, respectively, are shown by  blue solid curves in the Figs. \ref{spectrum20} (a) $\&$ \ref{spectrum20}(b), whereas, to see the region-wise contributions to the total cross-section, the corresponding cumulative integrated curves, shown by the green dotted lines, are also given.\footnote{ The cumulative integrated curve is defined by $F(x) = \int\limits_{x_0}^{x}f(x^\prime)dx^\prime$, where the function $f(x)$ represents the given curve.}
The energy spectrum in Fig. \ref{spectrum20}(b) contains the electron energy  ranging from $\sim 2-185~\rm{MeV}$, showing the acceleration. Also, there is a mono-energetic peak found at  $E_f=169.2~\rm{MeV}$ (the corresponding angle is $\theta_e=0.685$) with energy spread $\Delta E/E =0.06\%$; the peak contributes to $48\%$ of the total cross-section. However, as we can see in Fig. \ref{spectrum20}(a), the corresponding angular peak at $\theta_e=0.685$ (as can be found using Fig. \ref{spectrum20}(c)) contributes only to $6.3\%$ of the total cross-section$-$showing a very poor collimation.
The beam quality index $Q$, as defined in Eq. (\ref{BQI}), is found to be $Q=3.58$.

\section{Sensitivity of the Spectrum on Plasma Density  and the Electron Distribution}
\label{sec:sec_5}
\noindent In this section, we illustrate the sensitivity of the final spectrum on the parameters: the laser frequency and the plasma density, and also on the choice of the initial electron distribution. For example, the spectrum quality, shown in Fig. \ref{spectrum20}, is readily enhanced with the following choice of parameters : initial frequency $\omega_i =16 \omega_p$, plasma density $n_e=5.53\times 10^{18}~ \rm{{cm}^{-3}}$ and temperature $T=50~ \rm{eV}$, which is shown in Fig. \ref{spectrum}. The angular and the corresponding energy spectrum are shown, respectively, in the Fig. \ref{spectrum}(a) $\&$ Fig. \ref{spectrum}(b).   The dominant angular peak is found at the angle $\theta_e=0.541$ with spread $\Delta\theta\sim 3.75~ \rm{mrad}$; the corresponding energy peak (which can be found using Fig. \ref{spectrum}(c)) is found at energy $E_f=109.4~ \rm{MeV}$, with an energy spread of $\Delta E/E=0.64\%$ which contributes to $58\%$ of the cross-section, hence is a collimated and mono-energetic behavior of the spectrum. The beam quality index is found to be $ Q=48.33$ which is significantly greater than that found for the spectrum in Fig. \ref{spectrum20}, i.e., $Q=3.58$. Also, we can compare this spectrum (after $N=10^8$) with the initial electron distribution (with $4\%$ of energy spread and uniform angular distribution); we notice a drastic change in the nature of the spectrum through the scatterings$-$ the mono-energetic behavior and collimation has significantly enhanced. 

\begin{figure}
   \includegraphics[clip,width=8 cm,height=11 cm]{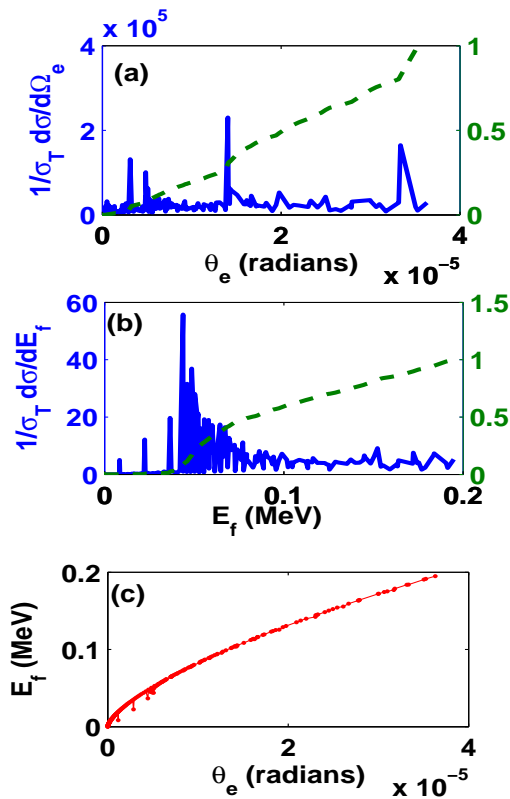}
   \caption{The electron spectrum with poor energy yield, at the density $n_e=5.53\times 10^{18}~ \rm{cm^{-3}}$ and temperature
   $T=50~\rm{eV}$, when the initial electron distribution is confined to the directions $\theta_i \in [1.1, 1.5]$ radians. }
   \label{spectrum_poor}
\end{figure}

With the above set of parameters, if we choose the direction of the electron momenta  $\theta_i \in [1.1,1.5]$, the 
 full spectrum is found to be within $\sim 0.04~ \rm{mrad}$, which indicates a   high collimation. However, the energy yield is very poor (the maximum acceleration is found only upto $0.2~\rm{MeV}$), as is shown in the Fig. \ref{spectrum_poor}. 

\begin{figure}
\includegraphics[clip,width=8 cm,height=11 cm]{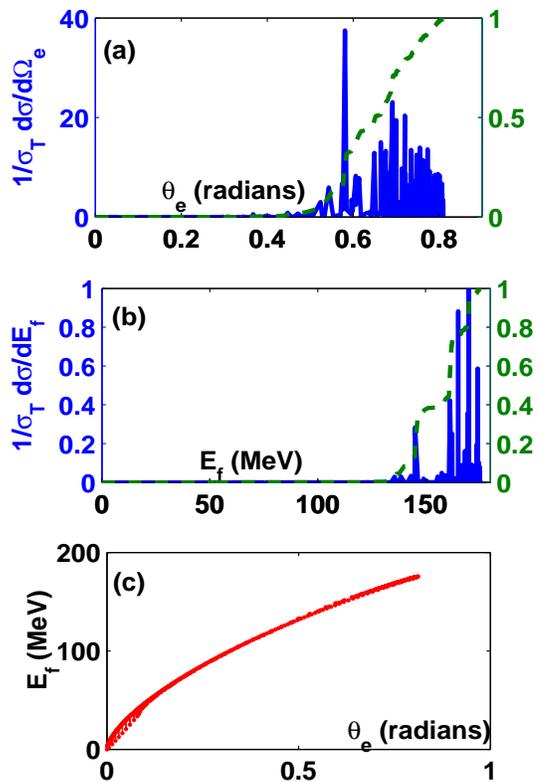}
\caption{The electron spectrum at density $n_e=8.53\times 10^{18}~ \rm{cm^{-3}}$ and temperature $T=50 ~\rm{eV}$,
when the initial electron distribution is confined to the directions $\theta_i \in [2.1, 2.5]$ radians. The beam quality index $Q=0.83$.}
\label{spectrum_12_2.46}
\end{figure}

The spectrum, after varying the plasma density to $n_e=8.53 \times 10^{18}~ \rm{cm^{-3}}$ and keeping the other parameters the same, i.e., $\omega_i =16 \omega_p,~T=50~\rm{eV}$ and $\theta_i \in [2.1, 2.5]$, is shown in Fig. \ref{spectrum_12_2.46}. The spectrum quality is found to deteriorate ($Q=0.83$)  as compared to the spectrum shown in Fig. \ref{spectrum}, however, the energy yield is better. 

\section{Discussion and Conclusion}
\label{sec:sec_6}
 \noindent We have demonstrated the acceleration of electron via multi-scattering mechanism $-$ the energies of the order of $O(100~\rm{MeV})$ within a centimeter are indeed possible to attain by electrons in a medium such as plasma. Also, we have demonstrated that a quality of the final spectrum is sensitive to the initial distribution, plasma parameters and initial photon frequency; the right choice of these can indeed lead a good quality of spectrum, such as in Fig. \ref{spectrum}. The mechanism behind the attainment of these collimated and mono-energetic peaks is the same as illustrated in I and II. That is, the spectrum (differential cross-section curve) attains its maxima near the resonance conditions: $\omega_i=\vec v_i.~\vec k_i$, and $\omega_f=\vec v_i.~\vec k_f$ which means that the phase velocity of the electron is equal to that of the either initial or scattered photon.\footnote{There are so many peaks corresponding to these conditions, however, for the quality of the spectrum, we concern to the dominant peak only.}  Also, the position of the dominant peaks (the energy and corresponding angle) has a complicated dependence on plasma parameters. The quality of the spectrum varies according to how well these resonance conditions are satisfied by the electron momenta (after $N=10^8$ scatterings) and the initial plasma frequency $\omega_i$. The collective nature of the plasma plays a vital role behind all these results. 
 
The dependence of the beam quality on the plasma density has been reported in the experiments \cite{faure, geddes, mangles, malka, joshi1, esarey}, especially, in Ref. \cite{faure}. However, the mechanism were relied on a non-linear wake-field, whereas the present analysis is valid only in linear regime. Also, we have not taken the Volkov states of the electron into account which may change the kinematics and dynamics.
The other lacuna which we have not considered is the radiation emission by the electron as it gains acceleration, which further, causes deceleration to the electron. After removing these lacunae we hope to match our results with the experiments. 

However, there are several advantages of this EFT approach: (i) The acceleration mechanism (through the interplay of kinematics and dynamics) is quite evident in this approach. (ii) It provides a good control over the acceleration and final beam quality (through resonance conditions). (iii) The incorporation of the quantum corrections are much easier in this approach, which might be useful for future laser plasma experiments and, which may not easily accessible by other methods.

\appendix
\section{The Expression for the Scattering Probability $ {\left| \bar{M}_{\alpha\alpha^\prime} \right|^2} $ used in the Eq. \ref{crosssection}}
\label{app:appendix_a}
 Using the Feynman rules displayed
 in Fig. 1 of Ref. \cite{ravi2}, the expression for the scattering probability can be obtained.  
 After performing the summation and averaging over the electron spins (first summing over the final electron spins and then averaging over the initial ones), the expression for the scattering probability is given by 
   \begin{eqnarray}
         {\left| \bar{M}_{\alpha\alpha^\prime} \right|^2} &=& \frac{x}{y} + \frac{y}{x} - 2 + 4 a_0\nonumber\\
         &+& xd_1 + yd_2 \nonumber\\
         &+& x^2\{a_1+ {\Delta_i}^2{b_1} + {\Delta_f}^2{c_1} + {\Delta_i}^2{\Delta_f}^2\}\nonumber\\
         &+& xy\{a_2+{\Delta_i}^2{b_2} + {\Delta_f}^2{c_2} \nonumber\\
         &+& {\Delta_i}^2{\Delta_f}^2(- 2 + 4 a_0)\} \nonumber\\
         &+&y^2\{a_3+ {\Delta_i}^2{b_1}+ {\Delta_f}^2{c_3} + {\Delta_i}^2{\Delta_f}^2\}\nonumber\\
       \label{prob}
       \end{eqnarray}
        The kinematical factors, $x = \frac{1}{s - m^2}$ and $y = \frac{ - 1}{t - m^2}$ where $s =(P_i+k_i)^2$ and $t=(P_i-k_f)^2$ are the standard Mandelstam variables and, ${\Delta}^2_{i,f} \equiv \omega^2_{i,f}-\vec k^2_{i,f} $. The coefficients $a_j,~b_j,~c_j(j=1,2,3)~ $\&$ ~d_j(j=1,2)$ are complicated functions of the photon polarizations (the polarization indices have been suppressed for brevity) and, the momentum variables $P_i,~ k_i~\& ~k_f$. They are listed in Tables A.1 $\&$ A.2 of Ref. \cite{ravi1} (see also Appendix-A of Ref. \cite{ravi2}).

\section{Kinematical restriction for the $LL$-scattering}
\label{app:appendix_b}
We shall demonstrate here that there are further kinematical restrictions on energy transfer to the electron,, apart from the dispersion relation constraints,  in  $LL$-scattering.  The photon scattering angle is given by (Eq. (\ref{theta_f})) 
 \begin{eqnarray}
 \cos(\theta_f-\phi)=\frac{C}{R}=\frac{R^2+k_f^2-P_f^2}{2R k_f}.
 \end{eqnarray}
leading to the bound
\begin {equation}
 \big|\frac{R^2+k_f^2-P_f^2}{2R k_f}\big |\leq 1.
\end{equation}
Observing that  $R, k_f>0$, we get 
\begin{eqnarray}
-2R k_f\leq R^2+k_f^2-P_f^2\leq 2R k_f.
\end{eqnarray}
which in turn leads, after some algebra, to the inequality  $P_f-k_f\leq R\leq P_f+k_f $.\\

Since  $k_i/P_i \approx 10^{-6}$, the expression for $R$ may be approximated as
\begin{eqnarray}
R=(P_i^2 +k_i^2+2 P_i k_i \cos\theta_i)^{\frac{1}{2}} \approx P_i + k_i \cos\theta_i
\end{eqnarray}
Furthermore, since $\Delta\omega/E\approx 10^{-6}$, we can approximate $\Delta P\equiv P_f-P_i$ as follows:
\begin{equation}
\Delta P\approx \frac{E_i\Delta \omega}{P_i};~ \Delta \omega \equiv \omega_i -\omega_f
\end{equation}
The above approximations imply finally that
\begin{eqnarray}
\frac{-k_f+\Delta P}{k_i}\leq \cos\theta_i\leq \frac{k_f+\Delta P}{k_i}.
\label{ineq_cos}
\end{eqnarray} 
which severely constrains the scattering angle, and hence the energy gain. We note that in making this argument, we use the condition 
 $\omega_f<\omega_i$ which also implies $k_f<k_i$ in the case of $LL$-scattering. 
In the case of $TL$-scattering, it can easily be checked that $k_f > k_i$, given that the initial photon frequency $\omega_i$ is not so large. Indeed, for the laser frequency ranges used in laser-plasma experiments, $k_f>k_i$ is usually followed. Also,  the value of $k_f$ is restricted by the constraint $\frac{\omega_f}{\omega_p} \in [1, 1.28]$. Hence, there is no corresponding bound for $\theta_i$, which allows large energy transfers. 

This work was supported by the Council of Scientific and Industrial
     Research, New Delhi, India {\it  (09/092(0345)/2004-EMR-I)}.

      \appendix

\end{document}